\documentclass{ifacconf}

\usepackage{graphicx}      
\usepackage{natbib}        
\usepackage{amsmath}
\usepackage{amssymb}
\usepackage{mathtools}
\usepackage{enumitem}
\usepackage{arydshln}

\newtheorem{defi}{Definition}
\newtheorem{assumption}{Assumption}
\newtheorem{remark}{Remark}

\newcommand*{\R}{\ensuremath{\mathbb{R}}}

\newcommand*{\Um}{\ensuremath{U_{-}}}
\newcommand*{\Xm}{\ensuremath{X_{-}}}
\newcommand*{\Xp}{\ensuremath{X_{+}}}

\newcommand*{\X}{\ensuremath{\mathcal{X}}}
\newcommand*{\U}{\ensuremath{\mathcal{U}}}
\newcommand*{\Sh}{\ensuremath{\mathcal{S}}}
\newcommand*{\D}{\ensuremath{\mathcal{D}}}
\newcommand*{\Ph}{\ensuremath{\mathcal{P}}}
\newcommand*{\Si}{\ensuremath{\Sigma}}
\newcommand*{\Sid}{\ensuremath{\Sigma_\mathcal{D}}}
\newcommand*{\Sip}{\ensuremath{\Sigma_\mathcal{P}}}
\newcommand*{\Sipk}{\ensuremath{\Sigma_\mathcal{P(K)}}}
\newcommand*{\Sidw}{\ensuremath{\Sigma_{\mathcal{DW}}}}

\newcommand*{\Kh}{\ensuremath{\mathcal{K}}}

\newcommand*{\Aj}{\ensuremath{A^{(j)}}}
\newcommand*{\Bj}{\ensuremath{B^{(j)}}}

\begin{document}
\begin{frontmatter}

\title{LMI-based Data-Driven Robust Model Predictive Control}


\author[TUD]{Hoang Hai Nguyen} 
\author[TUD]{Maurice Friedel} 
\author[TUD]{Rolf Findeisen}


\address[TUD]{Laboratory for Control and Cyber-Physical Systems, Technical University of Darmstadt, Germany. \newline
\tt\small\{hoang.nguyen, 
rolf.findeisen\}@iat.tu-darmstadt.de}


\begin{abstract}

Predictive control, which is based on a model of the system to compute the applied input optimizing the future system behavior, is by now widely used. If the nominal models are not given or are very uncertain,  data-driven model predictive control approaches can be employed, where the system model or input is directly obtained from past measured trajectories. Using a data informativity framework and Finsler's lemma, we propose a data-driven robust linear matrix inequality-based model predictive control scheme that considers input and state constraints for linear parameter-varying systems and Lur'e-type nonlinear systems.
Using these data, we formulate the problem as a semi-definite optimization problem, whose solution provides the matrix gain for the linear feedback, while the decisive variables are independent of the length of the measurement data. The designed controller stabilizes the closed-loop system asymptotically and guarantees constraint satisfaction. 
Numerical examples are conducted to illustrate the method.

\end{abstract}



\end{frontmatter}

\section{Introduction}

Model predictive control (MPC, see e.~g. \cite{rawlings}, \cite{findeisen2007}, \cite{lucia2016}) has become a popular control scheme thanks to the ability of efficiently handling constraints and performance criteria as well as the coherent implementation of multiple hierarchical layers.
The MPC scheme formulated as a semi-definite optimization problem in form of Linear Matrix Inequalities (LMIs) is often used for some typical classes of systems such as linear parameter-varying systems or Lur'e type systems, see for example, \cite{KOTHARE1996}, \cite{Bohm2009} and \cite{nguyen2018}.
The reason for this is the formulated optimization problem is convex and can be efficiently solved. 
Conventional MPC requires a model of the plant, which is usually obtained from first principle modelling or measured data via system identification. 
However, model development has been shown to be a critical and time-consuming step in MPC implementation (\cite{Rossiter2001}), while the performance of MPC elementally depends on the quality of the model.
An alternative approach is to design an MPC controller directly from measured data, without prior knowledge of an accurate model, which is often referred to as data-driven MPC. The learning-based approach, with the usage of Gaussian process (\cite{Hewing2020}, \cite{Maiworm2021}) or Reinforcement Learning (\cite{Zanon2021}), received increasing attention recently. 
The drawback of these approaches is that they imperil the advantage of MPC in handling constraints and providing theoretical guarantees since proving the desirable closed-loop properties is challenging.  

With regards to direct data-driven approaches, the fundamental result proposed by \cite{willems2005}, which is often referred as Willems et al. 's fundamental lemma, provides an answer on how to design controllers directly from data, where the system is implicitly represented via the Hankel matrix of measured trajectory. 
In short, the lemma states that all trajectories of an controllable LTI system can be represented by a finite set of its past trajectories, given that the past trajectories are generated by a sufficiently exciting inputs.
This idea has been investigated by \cite{persis2020}, where the stabilizing feedback gain is designed by formulating the problem as LMIs without considering performance and constraints guarantees, and \cite{Berberich2021}, \cite{Coulson2019b} where this idea is used to develop a data-driven MPC schemes. These approaches use persistently exciting data in control design; thus, the data have to be sufficiently rich enough to uniquely identify the system model within  a given model class.

Proposing the data informativity framework, the work \cite{Waarde2020} has shown that the corresponding informativity conditions on the data are often weaker than those for system identification in the sense that, in some cases, the data do not need to be sufficiently informative to uniquely identify the system. 
Aligned with this framework, the matrix Finsler's lemma provided in \cite{Waarde2021} (also see for \cite{Waarde2022}) provides a less conservative for constructing controllers from noisy data.
Another advantage of Finsler's lemma over the parameterizations approach based on Willems' fundamental lemma, which is commonly used, (such as in \cite{persis2020}, \cite{Berberich2021}) is that the decision variables are independent of the time horizon of the experiment as long as it is sufficiently long. 
This can play an important role in control design when working with big data sets.

In this paper, we propose a data-driven MPC scheme on the data informativity framework in form of LMIs, which designs an optimal feedback control gain from the data for systems subject to uncertainty in two cases: slowly varying linear systems and Lur'e systems with uncertain nonlinearity. 
This work can be considered as the corresponding data-driven results to the model-based results shown in \cite{KOTHARE1996}, \cite{Bohm2009}.
Since we base our scheme on data informativity and Finsler's lemma, our design method does not depend on the length of the experiments (such as in \cite{Berberich2021}), which is well suited for larger data sets.

The paper is structured as follows. 
Section 2  recapitulates the main ideas of the informativity framework for direct data-driven control and the matrix Finsler's lemma results.
We establish the LMI-based MPC scheme for the nominal case in Section 3.
In Section 4 and 5, we develop the results for the cases of slowly varying systems and Lur'e systems, respectively.
Section 6 presents numerical examples to illustrate the methods.
Finally, conclusions and outlook are given in Section 7.

\section{Data informativity for control design}
\label{sec2}

 We first recapitulate the main ideas of the \textit{informativity framework} for direct data-driven control proposed in \cite{Waarde2020}.
Let us consider an unknown system $\Sh$, which is contained in the model class $\Si$.
Furthermore, let us assume that we have access to a set of data set $\D$, which are generated by the system $\Sh$.
Given the data set $\D$, we define $\Sid \subseteq \Si$ as the set of all systems that could have generated the data.
The main concern of this paper is to design a controller for the system $\Sh$ from the data set $\D$ such that the closed-loop system, which consists of the system $\Sh$ and the controller, has a specific property.
Let $\Ph$ be a system-theoretic property and we denote the set of all systems in $\Si$ which have the property $\Ph$ as $\Sip$.
Given a controller $\Kh$, we denote the system-theoretic property $\Ph(\Kh)$ that is associated with the controller $\Kh$.
We only can achieve our control design objective if the chosen controller $\Kh$ guarantees that the closed-loop system between the controller and \textit{any} systems from the set $\Sid$ has the specified property. This leads to the definition of \textit{informativity for control}.
\begin{defi} \label{def1}
The data $\D$ are \textit{informative} for the property $\Ph(\cdot)$ if there  exists a controller $\Kh$ such that $\Sid \subseteq \Sipk$.
\end{defi}

In order to illustrate the framework, we consider a controllable and observable  discrete-time linear system of the form
\begin{equation}
   \quad x(k+1)=Ax(k)+Bu(k), \label{eq:plant}
\end{equation}
where $x \in \R^n, u \in \R^m$ are the states and control inputs of the system.
The real matrices $A$ and $B$ are assumed to be unknown.
Instead, we assume that we can collect the input and states data from system \eqref{eq:plant}, which are written in the matrices form
\begin{equation}
\begin{aligned} \label{eq:data}
    X &= [x(0) \quad  x(1) \quad \cdots x(T)],\\ 
    \Um &= [u(0) \quad u(1) \quad \cdots u(T-1)].\\
\end{aligned}
\end{equation}   
We denote the shifted versions of the states as follows.
\begin{equation} \label{eq:dataX}
\begin{aligned}
    \Xm &= [x(0) \quad  x(1) \quad \cdots x(T-1)],\\ 
    \Xp &= [x(1) \quad x(2) \quad \cdots x(T)].\\
\end{aligned}
\end{equation} 
In this case, $\Sigma$ is the model class of controllable and observable  discrete-time linear systems; the "true" system $\Sh$ is corresponding to a particular value of $(A,B)$ but we do not know. Then, we can define the data set $\D=(\Um,X)$. Consequently, the set $\Sid$ is equal to
\begin{equation} \label{eq:data_setD}
    \Sid:= \{(A,B) ~ | ~ \Xp=A\Xm + B\Um \}.
\end{equation}
If we aim to design a controller $\Kh$ in the form $u=Kx$ with $K\in \R^{m \times n}$ to stabilize the closed-loop system, then the property we want to achieve is $\Ph(\Kh):$~"state feedback $K$ yields a stable closed-loop system."
The corresponding set $\Sipk$ is equal to 
\begin{equation}
    \Sigma_{stab} := \{(A,B) ~ | ~ A+BK ~\textrm{is stable} \}.\footnote {For discrete systems, we say a matrix is stable when all of its eigenvalues lie inside the open unit disc.}
\end{equation}
In this case, the data $(\Um,X)$ are \textit{informative} for stabilization by state feedback if there exists  a $K$ such that $\Sid \subseteq \Si_{stab}$.

There are two problems that we need to answer.
The first one is whether the given data is "sufficient" to obtain a suitable controller from them, which leads to the \textit{informativity problem for control}.
\begin{prob}
Provide conditions on $\D$ under which there exists a controller $\Kh$ such that $\D$ are informative for the property $\Ph(\cdot)$.
\end{prob}
After the existence of the controller is ensured, the next problem is how to determine such controller, which is, in the framework, referred as the \textit{control design problem}.
\begin{prob}
Assume that the data $\D$ are informative for property $\Ph(\cdot)$, find a controller $\Kh$ such that $\Sid \subseteq \Sipk$.
\end{prob}

In this paper, we also use the matrix Finsler's lemma shown in \cite{Waarde2021}.
\begin{lem}(\cite{Waarde2021}) \label{lemma_waarde}
Consider symmetric matrices $M, \Xi \in \mathbb{R}^{(k+l) \times (k+l)}$, partitioned as
\begin{equation}
    M=
    \begin{bmatrix}
    M_{11} & \quad M_{12}\\
    M_{12}^\top & \quad M_{22}
    \end{bmatrix}
    ~\textrm{and}~
    \Xi=
    \begin{bmatrix}
    \Xi_{11} & \quad \Xi_{12}\\
    \Xi_{12}^\top & \quad \Xi_{22}
    \end{bmatrix}.
\end{equation}
Assume that (i) $M_{12}=0$ and $M_{22} \preceq 0$, (ii) $\Xi_{22} \preceq 0$ and $\Xi_{11}-\Xi_{12}\Xi_{22}^\dagger \Xi_{12}^\top=0$, (iii) $\exists G$ such that $M_{11} + G^\top M_{22} G \succ 0$ and $\Xi_{22}G=\Xi_{12}^\top$. Then, we have 
\begin{equation*}
    \begin{bmatrix}
    I\\
    Z
    \end{bmatrix}^\top
    M 
        \begin{bmatrix}
    I\\
    Z
    \end{bmatrix} \succeq 0 ~ \forall Z \in \mathbb{R}^{l \times k} \textrm{such that} 
        \begin{bmatrix}
    I\\
    Z
    \end{bmatrix}^\top
     \Xi
        \begin{bmatrix}
    I\\
    Z
    \end{bmatrix} = 0
\end{equation*}
if and only if there exists $\epsilon \in \mathbb{R}$ such that $M - \epsilon \Xi \succeq 0$.
\end{lem}

\section{Nominal LMI-based Data-Driven MPC}
\label{sec3}

In this section, we consider controlling system \eqref{eq:plant} subject to constraints $x \in \X$ and $u \in \U$ where the matrices $A$ and $B$ are unknown, but we have access to the data \eqref{eq:data}.
The constraint sets satisfy the following assumption. 
 \begin{assumption}\label{assumption_constraints}
 The constraint sets $\X$ and $\U$ can be described in the form
\begin{equation}
     \begin{aligned} \label{eq:constraintslinear}
     \X &= \{x \in \R^{n}: c_{i_x}x \leq 1, i_x=1,..., \bar n\}\\ 
     \U &= \{u \in \R^{m}: d_{i_u}u \leq 1, i_u=1,..., \bar m\},
 \end{aligned}  
\end{equation} 
 where $c_{i_x} \in \R^{1 \times n}$ and $d_{i_u} \in \R^{1\times m}$, and $\bar n, \bar m$ are the  numbers of states and inputs constraints, respectively. 
 It should be noted that \eqref{eq:constraintslinear} covers the case where $\X$ and $\U$ are polytopic.
\end{assumption}
We aim to find a feedback controller of the form
\begin{equation}\label{eq:control}
    u=Kx.
\end{equation}
The controller aims to minimize the infinite horizon cost functional
\begin{equation}\label{eq:cost_FunctionJ}
    J(x(\cdot),u(\cdot))= \sum_{k=0}^{\infty}x^\top (k) Q x(k) + u(k)^\top R u(k),
\end{equation}
where $Q \succ 0$ and $ R \succ 0$ are chosen weighting matrices.

Aligned with the Definition~\ref{def1}, we can define the data informativity for the case of LMI-based MPC for system \eqref{eq:plant}.
\begin{defi}
Given positive definite matrices $Q \ \in \R^{n \times n}$ and $R  \in \R^{m \times m}$ and initial state $x_0 \in \R^n$; and suppose that the data $(\Um,X)$ in \eqref{eq:data} are generated by \eqref{eq:plant}. 
Then, the data $(\Um,X)$ are called \textit{informative for designing the LMI-based data-driven model predictive controller} if there exists $K \in \R^{m \times n}$ such that the cost function \eqref{eq:cost_FunctionJ} is optimized and the constraints \eqref{eq:constraintslinear} are satisfied for all $(A,B) \in \Sid$ defined in \eqref{eq:data_setD}.
\end{defi}
It should be noted that the constraints \eqref{eq:constraintslinear} can be transformed into
\begin{align}
   \mathcal{W}=\{[x^\top~u^\top]^\top \in \R^{n+m}:c_i x + d_i u \leq 1\}, \label{eq:constraintsetW}
 \end{align}
 where $i=1,...,r$, with $r=\bar n+ \bar m$.
%
We need the following lemma for guaranteeing constraints satisfaction.
 \begin{lem}(\cite{boyd1994}) \label{lemma:eplipsoid}
 An ellipsoid
 \begin{equation} \label{eq:ellipsoidE}
     \mathcal{E}(\alpha)=\{x \in \R^{n}: x^\top P x \leq \alpha\}
 \end{equation}
 is contained in the set $\mathcal{W}=\{x \in \R^{n}: w_ix \leq 1, i=1,...,r\}$ if and only if 
 \begin{equation}
 w_i(\alpha P^{-1}) w_i ^\top \leq 1, i =1,...,r.    
 \end{equation}
\end{lem}
The following theorem establishes the results for the nominal case.
\begin{thm}\label{theorem1_nominal}
Let $x_0 \in \R^n$. The data $(\Um,X)$ are informative for designing the LMI-based data-driven model predictive controller if the following optimization problem is feasible: 
\begin{subequations} \label{eq:theorem1_prob}
	\begin{align}\nonumber
	&\min_{0 \prec N= N^\top \in \R^{n \times n},L\in \R^{m \times n},\alpha, \eta, \epsilon  >0}  ~\alpha \quad     \\ 
	\intertext{subject to}
	&\begin{bmatrix} \label{eq:theorem1_ellip}
    1 & \quad x_0^\top\\ 
    x_0& \quad N
    \end{bmatrix}  \succ 0, \\
    & \small \begin{bmatrix} \label{eq:theorem1_stab}
    N-\eta I & 0 & 0 & 0 & 0\\
    0 & 0 & 0 & N & 0 \\
    0 & 0 & 0 & L & 0 \\
    0 & N & L^\top & N & \Psi^\top \\
    0 & 0 & 0 & \Psi& \alpha I \\
    \end{bmatrix} +
    \epsilon \begin{bmatrix}
    \Xp \\
    -\Xm\\
    -\Um\\
    0\\
    0
    \end{bmatrix}
    \begin{bmatrix}
    \Xp \\
    -\Xm\\
    -\Um\\
    0\\
    0
    \end{bmatrix}^\top
    \succ 0, \\
      & \begin{bmatrix} \label{eq:theorem1_stab2}
      N    &    \quad \Psi^\top\\
      \Psi   &  \quad    \alpha I
     \end{bmatrix}   \succ 0,\\ 
    & \begin{bmatrix} \label{eq:theorem1_constraints}
      1    &    d_i L + c_i N\\
      (d_i L + c_i N)^\top   &     N
     \end{bmatrix}   \succ 0,\\ \nonumber
    & i=1,..., r,
	\end{align}
\end{subequations}
where $\Psi:= \hat Q N+ \hat R L$, $\hat Q = \begin{bmatrix}
Q^{1/2}\\
0
\end{bmatrix}$, $\hat R = \begin{bmatrix}
0\\
R^{1/2}
\end{bmatrix}$.
When \eqref{eq:theorem1_prob} is solved, $K:= L N^{-1}$ asymptotically stabilizes the closed-loop system and the input and states constraints \eqref{eq:constraintslinear} are satisfied. Also, $\alpha$ is the upper bound on the infinite horizon cost functional \eqref{eq:cost_FunctionJ}.
\end{thm}
\begin{pf}\\
Consider the Lyapunov function $V(k)=x(k)^\top P x(k)$ for the closed-loop system \eqref{eq:plant} with the linear feedback controller \eqref{eq:control}.  Let us consider the condition
\begin{equation} \label{eq:proof_Lyapunov}
    V(k+1)-V(k) < -x^\top (k) Q x(k) - u(k)^\top R u(k)\quad \forall k,
\end{equation}
holds, which is equivalent to
\begin{equation*}
    (A + BK)^T P (A+BK) - P < - ( \hat{Q} + \hat{R} K )^T ( \hat{Q} + \hat{R} K ),
\end{equation*}
with $\hat Q = \begin{bmatrix}
Q^{1/2}\\
0
\end{bmatrix}$, $\hat R = \begin{bmatrix}
0\\
R^{1/2}\end{bmatrix}$. Multiplying both sides with $N$, then changing variables $P=\alpha N^{-1}$, $K=LN^{-1}$ with $N=N^\top \succ 0$ and $\alpha >0$, we obtain
\begin{equation*}
\begin{aligned}
        N &- (AN + BL)^\top N^{-1} (AN + BL) -\frac{1}{\alpha}\Psi^\top \Psi> 0,
\end{aligned}
\end{equation*}
where $\Psi:= \hat Q N+ \hat R L$.
Using Schur complement leads to the following inequalities 
\begin{align*}
&\begin{bmatrix}
N- \frac{1}{\alpha}\Psi^\top \Psi  & \quad \quad (AN + BL)^\top \\
AN + BL & N
\end{bmatrix} \succ 0, \\
&N > 0,
\end{align*}
which results in the following inequalities
\begin{subequations}
\begin{align} \label{eq:proof1a1}
&N  - \frac{1}{\alpha}\Psi^\top \Psi> 0, \\ %
&N-(AN + BL)(N  - \frac{1}{\alpha}\Psi^\top \Psi )^{-1} (AN + BL)^\top \succ 0. \label{eq:proof1a}
\end{align}
\end{subequations}

Inequality \eqref{eq:proof1a1} is equivalent to condition \eqref{eq:theorem1_stab2}, while inequality \eqref{eq:proof1a} is equivalent to
\begin{align} \label{eq:proof1_f1}
\begin{bmatrix}
I \\ A^T \\ B^T
\end{bmatrix}^T
\begin{bmatrix}
N & 0 \\ 0 & -\begin{bmatrix}
N \\ L
\end{bmatrix} (N  - \frac{1}{\alpha}\Psi^\top \Psi)^{-1} \begin{bmatrix} N \\ L \end{bmatrix}^\top
\end{bmatrix}
\begin{bmatrix}
I \\ A^T \\ B^T
\end{bmatrix} \succ 0.
\end{align}
Consider the data $\Sid$ in \eqref{eq:data_setD}, similar to \cite{Waarde2021}, \cite{Waarde2022}, we know  $(A,B) \in \Sid$ if and only if $(A,B)$ satisfies 
%
\begin{equation} \label{eq:proof1_f2}
    \begin{bmatrix}
I \\ A^T \\ B^T
\end{bmatrix}^T
\begin{bmatrix}
 \Xp \\
 -\Xm \\
 -\Um
\end{bmatrix}
\begin{bmatrix}
 \Xp \\
 -\Xm \\
 -\Um
\end{bmatrix}^\top
\begin{bmatrix}
I \\ A^T \\ B^T
\end{bmatrix} = 0.
\end{equation}
Compare all assumptions of Lemma~\ref{lemma_waarde} for the two matrices
\begin{equation*}\small
    M=
    \left[
    \begin{array}{c;{2pt/2pt}c}
        M_{11} & M_{12} \\ \hdashline[2pt/2pt] \\[0.5 pt]
        M_{12}^\top & M_{22} 
    \end{array}
\right]
=
  \left[
    \begin{array}{c;{2pt/2pt}c}
        N & 0 \\ \hdashline[2pt/2pt] \\[0.5 pt]
        0 &  -\begin{bmatrix}
N \\ L
\end{bmatrix} (N  - \frac{1}{\alpha}\Psi^\top \Psi)^{-1} \begin{bmatrix} N \\ L \end{bmatrix}^\top

    \end{array}
    \right],
\end{equation*}
\begin{equation*}\small
    \Xi=
    \left[
    \begin{array}{c;{2pt/2pt}c}
        \Xi_{11} & \Xi_{12} \\  
        \hdashline[2pt/2pt]\\[0.5 pt]
          \Xi_{12}^\top & \Xi_{22} 
    \end{array}
\right]
= -
\left[
    \begin{array}{c}
        \Xp \\ \hdashline[2pt/2pt] \\[0.5 pt]
        -\Xm\\
        -\Um
    \end{array}
\right]
\left[
    \begin{array}{c}
        \Xp \\ \hdashline[2pt/2pt] \\[0.5 pt]
         -\Xm\\
        -\Um
    \end{array}
\right]^\top,
\end{equation*}
(i) holds with by the condition of \eqref{eq:theorem1_stab2}, (ii) and (iii) are satisfied by the reasons showed in Theorem~2 from \cite{Waarde2021}, so all assumptions are satisfied.
By applying Lemma~\ref{lemma_waarde}, we can obtain \eqref{eq:theorem1_stab} by using Schur complement. Since \eqref{eq:theorem1_stab} holds then \eqref{eq:proof_Lyapunov} holds, which means the system is asymptotically stable.

Moreover, summing up the two sides of \eqref{eq:proof_Lyapunov} with $k$ from $0$ to $\infty$, we obtain
\begin{equation}
    \sum_{k=0}^{\infty}x^\top (k) Q x(k) + u(k)^\top R u(k) < x_0 ^\top P x_0-V(\infty).
\end{equation}
Since the system is asymptotically stable, $V(\infty) \rightarrow 0$.
By using Schur complement, inequality \eqref{eq:theorem1_ellip} is 
\begin{equation*}
     x_0^\top P x_0 < \alpha.
\end{equation*}
Therefore, $\alpha$ is the upper-bound of the cost functional \eqref{eq:cost_FunctionJ}. 

For constraint satisfaction, the idea is to fit the ellipsoid $\mathcal{E}(\alpha)$ inside the set $\mathcal{W}$ created by constraints by using Lemma~\ref{lemma:eplipsoid}.
Condition \eqref{eq:theorem1_constraints} leads to 
\begin{equation}
    (c_i + d_i K) \alpha P^{-1} (c_i + d_i K)^\top \leq 1,
\end{equation}
which satisfies the condition of Lemma~\ref{lemma:eplipsoid}. Therefore, the constraints are satisfied.\hfill $\blacksquare$
\end{pf}
\begin{remark}
The decisive variables $N$ and $L$ as well as the constraints \eqref{eq:theorem1_constraints} in optimization problem \eqref{eq:theorem1_prob} do not depend on the length $T$ of the experiment data, which is advantage in comparison with works such as \cite{persis2020}, \cite{Berberich2021}, where the variable often has the size $T \times n$. On the other hand, the approach to guarantee constraint satisfaction using ellipsoids can introduce conservatism. 
\end{remark}
\begin{remark}
The optimization problem \eqref{eq:theorem1_prob} can be solved offline once to obtain the gain $K$, or it can be solved online similar to \cite{Bohm2009}.
When solving the optimization problem online, the data can be updated by updating only the second term of the left-hand side of \eqref{eq:theorem1_stab2}.
\end{remark}

\section{Robust Data-Driven MPC for Linear parameter-varying systems}
\label{sec4}

The result for the nominal case in the previous section can be extended naturally to the case where the matrices $A$ and $B$ have bounded uncertainties or they vary inside some convex sets.
However, since we do not neither know the values of matrices $A$ and $B$ nor the uncertainties, we need to assume that we are somehow able to collect data that sufficiently reflect the uncertainties.

In this section, we consider the polytopic plant described by
\begin{equation}\label{eq:plant_robust}
    x(k+1)=\Aj x(k)+\Bj u(k), \quad [\Aj~ \Bj] \in \Omega, 
\end{equation}
where the set $\Omega$ is a polytope 
\begin{equation} \label{eq:convexhull}
    \Omega= \textrm{Co}\{[A_1~B_1], [A_2~B_2], ..., [A_\zeta~B_\zeta]\},
\end{equation}
and $\textrm{Co}$ refers to the convex hull.
In other words, if $[A\quad B] \in \Omega$, then for some $\lambda_j>0$ such that $\sum_{j=1}^{\zeta} \lambda_j=1$, we have 
\begin{equation*}
    [A~B]=\sum_{j=1}^{\zeta} \lambda_j[A_j~B_j].
\end{equation*}
Polytopic system can be used to approximate systems working in different operating points, linear time-varying systems, linear systems with multiplicative uncertainty and nonlinear systems whose the Jacobian lies inside the polytope $\Omega$ in \eqref{eq:convexhull} (\cite{KOTHARE1996}).
For the case of systems with different operating points and each (unknown) pair $[\Aj~\Bj]$ is the model of the system at each operating point, we can apply the method proposed in Section~\ref{sec3} at each operating point. 
For uncertain linear systems or linear parameter-varying systems, we may need to design a robust control law $u=Kx$ to stabilize the system as long as  $[A\quad B] \in \Omega$. 
However, since we do not know the vertices of the polytope $\Omega$, we need to assume that we are able to collect data that can provide us the information of them.
\begin{assumption}\label{assumption_const}
It is possible to gather informative input-state data $(\Um,X)^{(j)}$ (referred as data set $\mathcal{D}^{(j)}$) that capture the same dynamical characteristic. 
\end{assumption}
This assumption requires that for the considered parameter-varying linear system, at each vertex $j$, we can collect sufficient data for the controller design purpose.
It can be applicable for multi-mode systems, where the system works in a mode for a period of time before changing to other modes and then may change back to the first mode some time later. A robust approach is to design a controller that stabilizes the system in every mode.

The following theorem establishes the results for nominal systems.
\begin{thm}\label{eq:theorem2_varying}
Let $x_0 \in \R^n$. The data sets $(\Um,X)^{(j)}$ are informative for designing the LMI-based model predictive controller if the following optimization problem is feasible: 
\begin{subequations} \label{eq:theorem2_prob}
	\begin{align}
	&\min_{0 \prec N= N^\top \in \R^{n \times n},L\in \R^{m \times n},\alpha, \eta, \epsilon  >0}  ~\alpha  \quad     \\
	\intertext{subject to}
	&\begin{bmatrix} \label{eq:theorem2_ellip}
    1 & \quad x_0^\top\\ 
    x_0& \quad N
    \end{bmatrix}  \succ 0, \\
    & \small \begin{bmatrix} \label{eq:theorem2_stab}
    N-\eta I & 0 & 0 & 0 & 0\\
    0 & 0 & 0 & N & 0 \\
    0 & 0 & 0 & L & 0 \\
    0 & N & L^\top & N & \Psi^\top \\
    0 & 0 & 0 & \Psi& \alpha I \\
    \end{bmatrix} +
    \epsilon \begin{bmatrix}
    \Xp^{(j)} \\
    -\Xm^{(j)}\\
    -\Um^{(j)}\\
    0\\
    0
    \end{bmatrix}
    \begin{bmatrix}
    \Xp^{(j)} \\
    -\Xm^{(j)}\\
    -\Um^{(j)}\\
    0\\
    0
    \end{bmatrix}^\top
    \succ 0,\\ \nonumber
    &\forall j=1,...,\zeta,\\
      & \begin{bmatrix} \label{eq:theorem2_stab2}
      N    &    \quad \Psi^\top\\
      \Psi   &  \quad    \alpha I
     \end{bmatrix}   \succ 0,\\ \nonumber
    & \begin{bmatrix} \label{eq:theorem2_constraints}
      1    &    d_i L + c_i N\\
      (d_i L + c_i N)^\top   &     N
     \end{bmatrix}   \succ 0,\\ \nonumber
    & i=1,..., r,
	\end{align}
\end{subequations}
where $\Psi:= \hat Q N+ \hat R L$, $\hat Q = \begin{bmatrix}
Q^{1/2}\\
0
\end{bmatrix}$, $\hat R = \begin{bmatrix}
0\\
R^{1/2}
\end{bmatrix}$.
When \eqref{eq:theorem2_prob} is solved, $K:= L N^{-1}$ asymptotically stabilizes the closed-loop system and the input and state constraints \eqref{eq:constraintslinear} are satisfied. Also, $\alpha$ is the upper bound on the infinite horizon cost functional \eqref{eq:cost_FunctionJ}.
\end{thm}

\begin{pf}
The proof is similar to that of Theorem~\ref{theorem1_nominal}. The only difference is the condition \eqref{eq:theorem2_stab}, which needs to hold for all the vertex of the polytope \eqref{eq:convexhull}. Since $\Omega$ is convex, when the condition \eqref{eq:theorem2_stab} holds for all the vertices, they hold for all $[A\quad B] \in \Omega$, (see \cite{boyd1994}, \cite{KOTHARE1996}).\hfill $\blacksquare$
\end{pf}

\section{Robust Data-Driven MPC for uncertain Lur'e-type systems}
\label{sec4}

In this section, we establish the result to design a robust data-driven MPC for Lur'e systems, which are described by
\begin{equation} \label{eq:luresys}
    x(k+1)=Ax(k)+Bu(k)+E\gamma (H x(k)),
\end{equation}
where $x \in \R^n, u \in \R^m$. The matrices $A$, $B$ and $E$ (with appropriate dimensions) are unknown and $H$ is assumed to be known.
The unknown nonlinearity $\gamma: \mathbb{R} \rightarrow \mathbb{R}$ satisfies the sector bound condition
\begin{equation} \label{eq:sectorbound}
    \gamma(z) (\beta z - \gamma(z)) \geq 0,~\forall z \in \mathbb{R},
\end{equation}
where $\beta \in \mathbb{R}_+$ is a known constant. In other words, $\gamma(\cdot)$ lies in the sector-bound $[0,\beta]$.
The goal of the control task is to minimize the cost function \eqref{eq:cost_FunctionJ} while guaranteeing that the states $x$ and inputs $u$ satisfy the constraints \eqref{eq:constraintslinear} in Assumption~\ref{assumption_constraints}.
For the data-driven approach, we assume that $A, B, E$ are unknown and we only can access the measurements of $X$ and $\Um$ as in \eqref{eq:data} as well as 
\begin{equation}
    W_{-}=[\gamma(Hx(0)) \quad \gamma(Hx(1)) \quad \dots \quad \gamma(Hx(T-1))].
\end{equation}
If $X$ and $X_{-}$ are defined as in \eqref{eq:dataX}, then we define the sets of all systems $(A,B,E)$ that can generate such data as
\begin{equation} \label{eq:datalure}
    \Sidw=\{(A,B,E) ~|~\Xp=A\Xm + B\Um + E W_{-}\}
\end{equation}
\begin{thm}\label{eq:theorem3_lure}
Let $x_0 \in \R^n$. The data $(\Um,X,W_{-})$ are informative for designing the LMI-based model predictive controller if the following optimization problem is feasible: 
\begin{subequations} \label{eq:theorem3_prob}
	\begin{align}
	&\min_{0 \prec N= N^\top \in \R^{n \times n},L\in \R^{m \times n},\alpha, \eta, \epsilon >0}  ~\alpha  \quad     \\
	\intertext{subject to}
	&\begin{bmatrix} \label{eq:theorem3_ellip}
    1 & \quad x_0^\top\\ 
    x_0& \quad N
    \end{bmatrix}  \succ 0, \\
    & \tiny \begin{bmatrix} \label{eq:theorem3_stab}
    N-\eta I & 0 & 0 & 0 & 0 & 0 &0\\
    0 & 0 & 0& 0 & 0 & N & 0 \\
    0 & 0 & 0 & 0 & 0 & L & 0 \\
    0 & 0 & 0 & 0 & \alpha I & 0 & 0 \\
    0 & 0 & 0 & \alpha I & \alpha I & -\dfrac{1}{2}\beta H N & 0 \\
    0 & N & L^\top & 0 & -\dfrac{1}{2} N H^\top \beta^\top & N & \Psi^\top \\
    0 & 0 & 0 & 0 & 0 & \Psi & \alpha I \\
    \end{bmatrix} +
    \epsilon \begin{bmatrix}
    \Xp \\
    -\Xm\\
    -\Um\\
    - W_{-}\\
    0\\
    0 \\
    0
    \end{bmatrix}
    \begin{bmatrix}
    \Xp \\
    -\Xm\\
    -\Um\\
    - W_{-}\\
    0\\
    0\\
    0
    \end{bmatrix}^\top \\ \nonumber
    & \quad \quad \quad \quad \succ 0, \\
     %
    &\begin{bmatrix} \label{eq:theorem3_stab2}
 N & -  \frac{1}{2}N H^\top \beta^\top   & \Psi^\top \\
  -  \frac{1}{2} \beta H N &  \alpha I & 0 \\    
\Psi & 0 & \alpha I 
 \end{bmatrix} \succ 0 \\
    & \begin{bmatrix} \label{eq:theorem3_constraints}
      1    &    d_i L + c_i N\\
      (d_i L + c_i N)^\top   &     N
     \end{bmatrix}   \succ 0,\\ \nonumber
    & i=1,..., r,
	\end{align}
\end{subequations}
where $\Psi:= \hat Q N+ \hat R L$, $\hat Q = \begin{bmatrix}
Q^{1/2}\\
0
\end{bmatrix}$, $\hat R = \begin{bmatrix}
0\\
R^{1/2}
\end{bmatrix}$.
When \eqref{eq:theorem1_prob} is solved, $K:= L N^{-1}$ asymptotically stabilizes the closed-loop system and the input and states constraints \eqref{eq:constraintslinear} are satisfied. Also, $\alpha$ is the upper bound on the infinite horizon cost functional \eqref{eq:cost_FunctionJ}.
\end{thm}
\begin{pf}\\
The sector-bound condition \eqref{eq:sectorbound} can be written in the quadratic form
\begin{equation} \label{eq:sectorbound2}
 \begin{bmatrix}
 x \\ z 
 \end{bmatrix}^\top
 \begin{bmatrix}
 0 & \frac{1}{2}H^\top \beta^\top \\
 \frac{1}{2} \beta H & - I
 \end{bmatrix}
 \begin{bmatrix}
 x \\ z
 \end{bmatrix} \succ 0.
 \end{equation}
Consider the Lyapunov function $V(k)=x(k)^\top P x(k)$ for the closed-loop system \eqref{eq:luresys} with the linear feedback controller \eqref{eq:control}.  We need to ensure the condition
\begin{equation} \label{eq:proof_Lyapunovlure}
    V(k+1)-V(k) < -x^\top (k) Q x(k) - u(k)^\top R u(k)\quad \forall k,
\end{equation}
holds when the sector-bound condition \eqref{eq:sectorbound2} holds.
Using the S-Lemma, we have 
 \begin{align*}
 \begin{bmatrix}
 P-A_K^\top P A_K - \Psi_K^\top \Psi_K & \quad \quad -A_K^\top P E -  \frac{1}{2}H^\top \beta^\top \\
 -E^\top P A_K - \frac{1}{2} \beta H &  I -E^\top P E
 \end{bmatrix} \succ 0,
 \end{align*}
where $A_K=A+BK$ and $\Psi_K = \hat{Q} + \hat{R} K$ . Changing variables $P=\alpha N^{-1}$ and $L=KN$, then multiplying both sides with $\textrm{diag}(N,I)$, where $N=N^\top \succ 0$, $\alpha >0$, we have
\begin{equation*} \small
 \begin{bmatrix}
 \alpha N -\alpha A_{NL}^\top N^{-1} A_{NL} - \Psi^\top \Psi & -\alpha A_{NL}^\top N^{-1} E -  \frac{1}{2} N H^\top \beta^\top  \\
 - \alpha E^\top N^{-1} A_{NL} -  \frac{1}{2}\beta H  N &  I -\alpha E^\top N^{-1} E
 \end{bmatrix}
 \succ 0.
 \end{equation*}
where $A_{NL}=AN+BL$. We can separate the matrix into two terms and then apply the Schur complement twice, which is similar to the techniques used in \cite{Waarde2021}. Then, we can obtain
%
%
 \begin{equation} \label{eq:lureprf1}
 \begin{bmatrix}
 I^\top \\
 A ^\top\\
 B ^\top\\
 E^\top
 \end{bmatrix}^\top
\Theta
 \begin{bmatrix}
 I^\top \\
 A ^\top\\
 B ^\top\\
 E^\top
 \end{bmatrix} \succ 0,  \\ 
\end{equation}
where
\begin{equation*}\small
    \Theta=
 \begin{bmatrix}
 N & 0 \\
 0 & -\begin{bmatrix}
 N  & 0 \\
 L & 0 \\
 0 & I
 \end{bmatrix} \alpha
 \begin{bmatrix}
\alpha N - \Psi^\top \Psi& -  \frac{1}{2} N H^\top \beta^\top \\
-  \frac{1}{2}\beta H N & I
\end{bmatrix}^{-1}
 \begin{bmatrix}
 N  & 0 \\
 L & 0 \\
 0 & I
 \end{bmatrix}^\top
 \end{bmatrix}    
\end{equation*}
From \eqref{eq:datalure}, $(A,B,E) \in \Sidw$ when
 \begin{equation}\label{eq:lureprf2}
 \begin{bmatrix}
 I^\top \\
 A ^\top\\
 B ^\top\\
 E^\top
 \end{bmatrix}^\top
\begin{bmatrix}
\Xp\\
-\Xm\\
-\Um\\
-W_{-}
\end{bmatrix}
\begin{bmatrix}
\Xp\\
-\Xm\\
-\Um\\
-W_{-}
\end{bmatrix}^\top
 \begin{bmatrix}
 I^\top \\
 A ^\top\\
 B ^\top\\
 E^\top
 \end{bmatrix} = 0,  \\ 
\end{equation}

Similar to Theorem~\ref{theorem1_nominal}, by applying Lemma~\ref{lemma_waarde} for \eqref{eq:lureprf1} and \eqref{eq:lureprf2} (and note that the condition $M_{22} \leq 0$ is guaranteed by \eqref{eq:theorem3_stab2}), we can obtain \eqref{eq:theorem3_stab}. Since \eqref{eq:theorem3_stab} holds then \eqref{eq:proof_Lyapunovlure} holds, which means the system is asymptotically stable.
The constraints satisfaction and the bound of $\alpha$ is the same as in Theorem~\eqref{theorem1_nominal} by using condtions \eqref{eq:theorem3_ellip} and \eqref{eq:theorem3_constraints}. \hfill $\blacksquare$

\end{pf}

\section{Numerical Examples}
\label{sec5}

We illustrate the proposed approach via two numerical examples.
The optimization problems in both examples are solved by using CVX toolbox (\cite{cvx1,cvx2}) with the solver SDPT3 (\cite{spdt3a}).
\subsection{Linear time-varying example} \label{sec61}
\begin{figure}[h]
\begin{center}
\includegraphics[width=8.4cm]{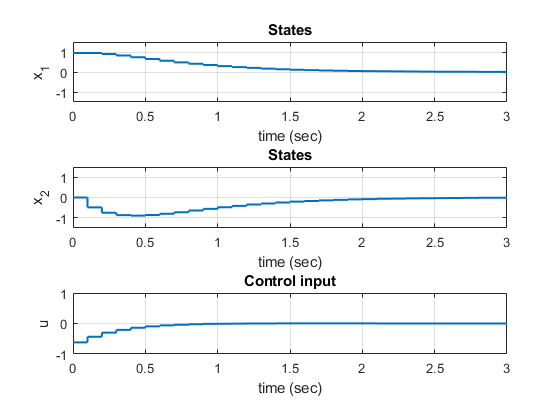}    
\caption{Simulation results for the example in Section~\ref{sec61}} 
\label{fig:closedloopresponsepolytop}
\end{center}
\end{figure}

We consider a model of angular positioning system used in \cite{KOTHARE1996}
\begin{equation*}\label{eq:plant_robust}
\begin{aligned}
    x(k+1)= & A(k) x(k)+B(k) u(k)\\
          = & 
          \begin{bmatrix}
          1 & 0.1\\
          0 & 1 - 0.1 \delta (k)
          \end{bmatrix} x(k)
          +
          \begin{bmatrix}
          0 \\
          0.1 \kappa
          \end{bmatrix} u(k)
\end{aligned}
\end{equation*}
where $\kappa=7.87$ and $ 0.1 \leq \delta (k)  \leq 10$.
Therefore, $A(k) \in \Omega= \textrm{Co}\{A_1, A_2\}$ where
\begin{equation}
    A_1=
    \begin{bmatrix}
    1 & \quad 0.1 \\
    0 & \quad 0.99 
    \end{bmatrix}, \quad
    A_2=
        \begin{bmatrix}
    1 & \quad 0.1 \\
    0 & \quad 0 
    \end{bmatrix}
\end{equation}
A sampling time of 0.1 seconds and the first-order Euler approximation were used to obtain the above model.
We consider the initial condition $x_0 = [0.95,0]^T$.
The weighting matrices are chosen as $Q = I$ and $R = 0.01$.
 The following input constraint need to be satisfied
 \[u(k)\in[-1,1].\]
We generate two random input sequences, each of length $T = 10$, and then applying them to the system to collect two data set $\mathcal{D}^{(1)}$ and $\mathcal{D}^{(2)}$ corresponding to each vertex, respectively.
Solving \eqref{eq:theorem2_prob}, we obtained
\begin{equation*}
    K=
    \begin{bmatrix}
    -0.6489 & -0.3809 
    \end{bmatrix}.
\end{equation*}
For the simulation results a system within the polytope was used with the system matrix $A_3 = 0.85 A_1 + 0.15 A_2$.
The simulation results are shown in Fig.~\ref{fig:closedloopresponsepolytop}, where the system is asymptotically stabilized and all constraints are satisfied.

\begin{figure}
\begin{center}
\label{fig:bifurcation}
\end{center}
\end{figure}

\subsection{Lur'e type systems} \label{sec62}
\begin{figure}[h]
 \centering
 \includegraphics[width=0.7\columnwidth]{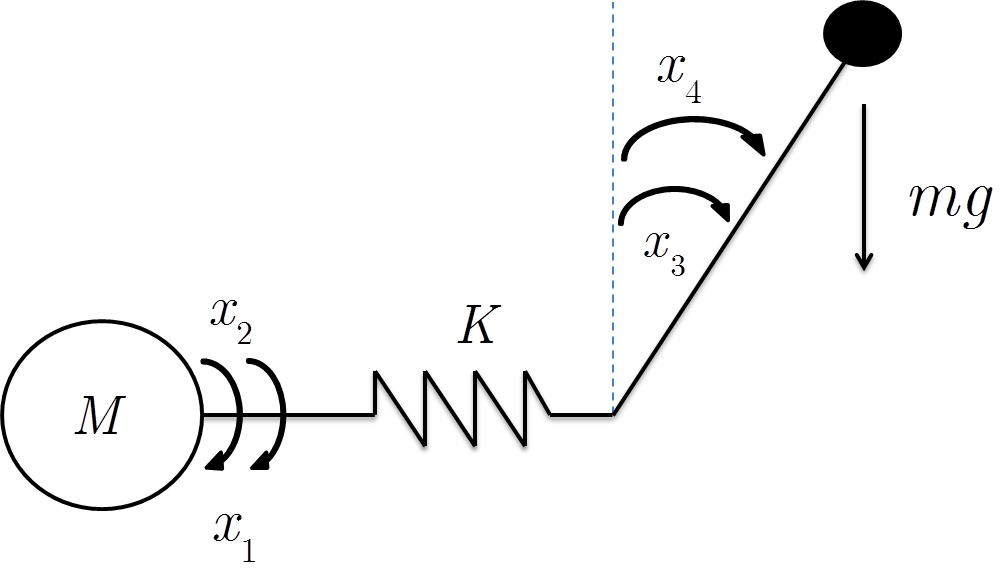}
    \caption{Graphical illustration of the flexible link robotic arm.}
    \label{fig:robotarm}
\end{figure}

We consider the control of a flexible robotic arm (Fig.~\ref{fig:robotarm}), described by 
\begin{equation}
\begin{aligned}
    x(k+1) &=Ax(k) + B u(k) + E (\gamma(z(k))),\\
    z(k) &= Hx(k)
\end{aligned}
\end{equation}
where
\begin{equation*}
 \begin{array}{l}
  A=\begin{bmatrix}
 1 & 0.02 & 0 & 0 \\
 -0.972 & 0.975 & 0.972 & 0\\
 0 & 0 & 1 & 0.02\\
 0.39 & 0 & -0.334 & 1 \\
 \end{bmatrix},~~~
 B=
 \begin{bmatrix}
 0\\
 0.432\\
 0\\
 0
 \end{bmatrix},\\
 E^T= \begin{bmatrix}
 0 & 0 & 0 & -0.0666
 \end{bmatrix},~~~
 H= \begin{bmatrix}
 0 & 0 & 1 & 0
 \end{bmatrix}.
  \end{array}
 \end{equation*} 
 A sampling time of 0.02 seconds and the first-order Euler approximation were used to obtain the above model.
 The considered nonlinear function $\gamma$ takes the form
 \begin{equation*}
 \gamma(z)=\sin(z) + z.
 \end{equation*}
 The sector-bound condition holds for all $\beta \geq 2$.
 We use the value $\beta = 2$ in the simulation.
 The following input and state constraints need to be satisfied
 \[u(k)\in[-2,2],\quad x_1(k),\,x_3(k)\in[-\frac{\pi}{2},\frac{\pi}{2}] \quad \forall k\geq 0.\]
We consider the initial condition $x_0 = [1.1,0.2,0,0]^T$.
The weighting matrices are chosen as
$Q = 0.1 \text{diag}(1, 0.1, 1, 0.1)$ and $R = 0.1$.
We generate the data by applying random input sequence of length $T = 50$.

Solving \eqref{eq:theorem3_prob}, we obtained
\begin{equation*}
    K=
    \begin{bmatrix}
    -1.0342  & -0.1949 &  -0.4329 &  -0.2236
    \end{bmatrix}.
\end{equation*}

\begin{figure}[h]
\begin{center}
\includegraphics[width=8.4cm]{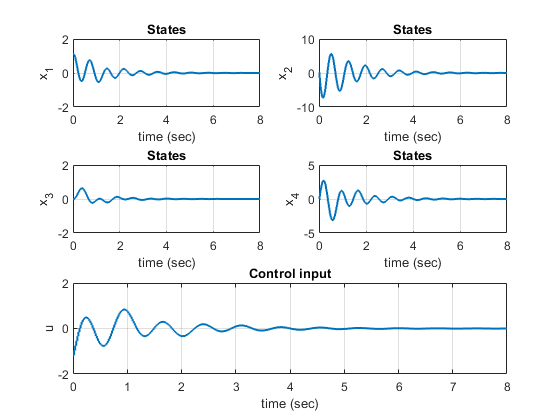}    
\caption{Simulation results for the example in Section~\ref{sec62}} 
\label{fig:closedloopresponselure}
\end{center}
\end{figure}

The simulation results are shown in Fig.~\ref{fig:closedloopresponselure}, where the system is asymptotically stabilized and all constraints are satisfied.

\section{Conclusions and Outlook}
\label{sec6}

In this paper, we present a data driven predicted control scheme to design controllers under state and input constraints for two classes of systems: (i) linear systems with multiplicative uncertainty (such as slowly varying linear systems or periodic systems) and (ii) Lur'e systems with sector-bounded nonlinearity. 
We developed our results based on the data informativity framework and Finsler’s lemma, and established the conditions in form of LMIs, which can be solved effectively with available toolboxes. 
The advantage of this approach in comparison with other data-driven approaches based on Willems' fundamental lemma is that decisive variables are independent of the length of the available experiment data.
The designed controller is proved to asymptotically stabilize the closed-loop system and guarantee constraints satisfaction. 
The method has been illustrated via two simulation examples.
For future work, we aim to extend the approach to the cases in which the system is affected by additive bounded noise and the different classes of nonlinearity.



\bibliography{ifacconf}             
                                                   







\end{document}